\documentclass[aps, prd, nofootinbib,12pt]{revtex4}

\usepackage[utf8]{inputenc}
\usepackage[T1]{fontenc}
\usepackage{amsmath,amssymb}
\usepackage{hyperref}
\usepackage{graphicx}
\usepackage{subfigure}

\usepackage[utf8]{inputenc}
\usepackage[T1]{fontenc}
\usepackage{amsmath,amssymb}

\DeclareMathOperator{\sgn}{sgn}

\begin{document}

\title{Primordial tensor modes of the early Universe}


\author{Florencia Ben\'itez Mart\'inez}
\affiliation{Instituto de F\'{i}sica, Facultad de Ciencias, Igu\'a 4225, esq.\ Mataojo, Montevideo, Uruguay}
\author{Javier Olmedo}
\affiliation{Department of Physics and Astronomy, Louisiana State University, Baton Rouge, Louisiana 70803-4001, USA}

\begin{abstract}
We study cosmological tensor perturbations on a quantized background within the hybrid quantization approach. In particular, we consider a flat, homogeneous and isotropic spacetime and small tensor inhomogeneities on it. We truncate the action to second order in the perturbations. The dynamics is ruled by a homogeneous scalar constraint. We carry out a canonical transformation in the system where the Hamiltonian for the tensor perturbations takes a canonical form. The new tensor modes now admit a standard Fock quantization with a unitary dynamics. We then combine this representation with a generic quantum scheme for the homogeneous sector. We adopt a Born-Oppenheimer ansatz for the solutions to the constraint operator, previously employed to study the dynamics of scalar inhomogeneities. We analyze the approximations that allow us to recover, on the one hand, a Schr\"odinger equation similar to the one emerging in the dressed metric approach and, on the other hand, the ones necessary for the effective evolution equations of these primordial tensor modes within the hybrid approach to be valid. Finally, we consider loop quantum cosmology as an example where these quantization techniques can be applied and compare with other approaches.
\end{abstract}


\maketitle

\newpage

\section{Introduction}

The properties of the large scale structures of the cosmos that we observe today provide an indirect picture of the state of the early Universe once the physical laws ruling its evolution are provided. Nowadays, there are several missions, like COBE, WMAP, Planck or BICEP \cite{precision}, among others, dedicated to the study of the cosmic microwave background (CMB). One of the most important observations is that our Universe is homogeneous and isotropic at large scales, with small relative inhomogeneities. If the evolution is described by means of a Friedmann-Robertson-Walker (FRW) metric fulfilling Einstein's equations, these small inhomogeneities should have been present in the early Universe in order to generate the current structures at smaller scales that we observe today. One of the most attractive descriptions in these lines is provided by the paradigm of cosmological perturbation theory \cite{perturbations,bardeen,theorypertu,mukhanov}. However, in order to describe the properties and structures of the present Universe in a natural way, it requires additional considerations. For instance, the theory of cosmological perturbation together with inflation \cite{inflation} and the principles of quantum mechanics provide a simple and rather satisfactory explanation. 

It is remarkable that this paradigm is traditionally based on the semiclassical approximation where quantum field theory (QFT) in a curved background is applicable. Therefore, the quantum nature of gravity at the early stages of the evolution is mostly neglected. Nevertheless, it seems indeed a very good approximation. The Gaussian distribution of temperature anisotropies found in the CMB is naturally explained by means of a primordial power spectrum of fluctuations of the comoving curvature perturbations if they are, at the onset of inflation, on (or close to) the Bunch-Davis state \cite{bunchdavies} (the vacuum state in a de Sitter spacetime). However, as it was first announced by WMAP and later confirmed by Planck, there is some evidence about the existence of anomalies (suppression at large scales) in the spectrum of temperature anisotropies indicating that new physics of the early Universe could be involved in the process. There have been several proposals trying to give an explanation to this phenomenon. One possibility is that the integrated Sachs-Wolfe effect \cite{isw} produces a suppression instead of an enhancement (as it is commonly believed) of the low-multipole temperature anisotropies power spectrum. This explanation depends crucially on the late evolution of the Universe. However, we find more attractive alternative ideas like models producing a dipole modulation in the observed primordial spectrum \cite{modulation} or suppression at large scales \cite{suppression}. It is worth commenting that quantum geometry corrections are also able to produce either a suppression \cite{lqc1} or an enhancement \cite{AAN2,AM} of the primordial power spectrum at large scales, as well as modulations \cite{nongauss}.

Concretely, loop quantum cosmology (LQC) \cite{lqc-rev}, the quantization of symmetry reduced models based on loop quantum gravity (LQG) \cite{lqg} ideas, has emerged as a serious branch of research in quantum cosmology and the early physics of the Universe. One of the main features of this new paradigm is the replacement of the classical singularity by a quantum bounce \cite{aps} (usually called big bounce). It provides a natural extension of the early Universe through the Planck era, where quantum gravity corrections are not negligible. It has been applied to the current inflationary isotropic models, mainly single-field inflation scenarios (a homogeneous minimally coupled scalar field with a quadratic potential) \cite{AAN2,AM,as,effective,effective1,hybrid,hybrid1}. Under certain semiclassical assumptions, the fine-tuning in the initial values of the geometry and the matter content is naturally solved if one gives initial data at the big bounce \cite{as}. Moreover, the traditional picture of QFT in a curved classical background has been extended to the one of QFT in effective backgrounds that incorporate quantum geometry fluctuations \cite{AAN2,AAN1}, the so-called dressed metric approach. 

In spite of the fact that all these developments provide a deeper understanding of the physics of the early Universe, they are not able yet to give a definitive answer to the initial value problem for the perturbations. Let us recall that in traditional inflation it is enough to assume that the perturbations are at (or close to) the Bunch-Davies vacuum at the onset of inflation. This is where the prediction capability of this formalism resides. Besides, in the case of LQC models it has been possible to find several candidates as initial vacuum states compatible with observations. For instance, in Ref. \cite{AAN2} the authors studied scalar and tensor perturbations initially at a fourth-order adiabatic state \cite{adiab}. If the number of $e$-foldings is enough, the primordial power spectrum seems to be compatible with current observations, but with a small enhancement at the large scales observed today. It seems also possible to choose among this family of adiabatic states a subset producing instead a suppression \cite{lqc1}. Besides, in Ref. \cite{AAN3}, it was proven that there exists a unique fourth-order adiabatic state at a given initial time (close to the bounce) where the stress-energy tensor of the perturbations vanishes. More remarkable it is the fact that LQC provides potential predictions. For instance, it seems that for those adiabatic states where the primordial power spectrum is suppressed \cite{lqc1}, LQC predicts a lower electric-electric (E-E) correlation function for large scales with respect to the one in standard inflationary cosmology with the integrated Sachs-Wolfe effect \cite{isw}. In addition, it is able to modify the consistency relation $r\approx -8n_t$ between the tensor-to-scalar ratio $r$ and the tensor spectral index $n_t$ of the standard inflationary scenario.

Let us comment that, in addition to the previous apparent freedom in the choice of initial vacuum state, there also exist several prescriptions in LQC for the quantization of cosmological perturbations on these inflationary scenarios. Within the different quantization approaches for cosmological perturbations in LQC, in this manuscript, we will focus on the hybrid quantization approach \cite{hybrid,hybrid1}. It was originally suggested in Ref. \cite{gowdy} as a quantization prescription of a gravitational system that admits a regime between full quantum gravity and the semiclassical approximation of QFT in curved spacetimes. One then combines a Fock representation for the linear inhomogeneities with a polymeric quantization for the remaining degrees of freedom. Its application to cosmological perturbations has been considered in Refs. \cite{hybrid,hybrid1}. This approach, contrary to the dressed metric approach, keeps backreaction contributions of the perturbations to second order in the action. In order for this truncation be consistent, the backreaction should be small on solutions. We must recall that it is a natural requirement in cosmological perturbation theory (at the level of truncation considered here), but keep in mind that the hybrid approach is valid in nonperturbative scenarios \cite{gowdy}. Besides, this hybrid quantization approach for cosmological perturbations admits a fully covariant description \cite{muk-hyb} for scalar inhomogeneities, but with strong differences with respect to the deformed algebra approach of Refs. \cite{effective,effective1}. In addition, it has been partially confronted with observations \cite{hyb-obs}. The hybrid approach is based on previous proposals \cite{hybrid,hybrid1} where different gauge fixings were adopted classically. At the quantum level, the perturbative approximation suggests quantization prescriptions that turn out to affect the semiclassical evolution of the perturbations with respect to other proposals \cite{AAN2,AM} and possibly modify qualitatively the final outcomes. In addition, this hybrid approach admits approximate solutions to the quantum scalar constraint where the dressed metric approach \cite{lqc1,AAN2,AM,AAN1} can be naturally accommodated as well as provides well-defined effective equations of motion for the perturbations with a suitable ultraviolet limit. Its physical consequences are still under investigation, but our preliminary analysis provides physical predictions compatible with the ones of Refs. \cite{lqc1,AAN2,AM} for the power spectrum of scalar perturbations. 

Since this formalism has been applied only to scalar perturbations, due to the fact that they are more interesting from the technical and physical point of view, in this manuscript, we will extend it in order to include tensor perturbations. Furthermore, this extension will allow us to analyze the main predictions of the hybrid quantization approach with respect to the power spectrum of primordial tensor modes and the relation between the tensor spectral index $n_t$ and the tensor-to-scalar ratio $r$. The high precision missions planned for the future will be focused on the study of the polarization of the CMB which would give an indirect measurement of primordial gravitational waves. So, it seems essential to incorporate tensor perturbations within this formalism and provide the effective equations of motion in order to study the predictions of the model. On the other hand, the simple form of the Hamiltonian for tensor modes with respect to the one of the scalar perturbations provides one of the simplest arenas where we can study the full quantum dynamics of this class of systems. Let us recall that in the particular case where the scalar field is massless the dynamics of the background is well known \cite{aps}. Therefore, it will be possible to confront the different quantization approaches for perturbations in LQC beyond the approximation where the usual effective dynamics is valid. 

This paper is organized as follows. In Sec. \ref{sec:system}, we describe the classical setting to be studied in this work. We then set the hybrid quantization in Sec. \ref{sec:quantization}. In Sec. \ref{BOa}, we study approximate solutions to the quantum scalar constraint within a Born-Oppenheimer ansatz, where we analyze the approximations required in order to recover the dressed metric approach and the effective equations of motion. We then consider a loop quantization for the homogeneous sector as a particular example where all these results can be applied, and we compare with other approaches. We conclude with Sec. \ref{conclusions}. In addition, we include one Appendix.

\section{Classical FRW spacetimes with small inhomogeneities}\label{sec:system}

Let us start with the classical description of the model. We will mainly follow Refs. \cite{muk-hyb,hybrid,hybrid1}. We consider a flat FRW spacetime with compact $T^3$ topology minimally coupled to a scalar field $\Phi$ subject to a quadratic potential (due to its special interest in observations) of mass $m$. The full action of the model is the time integral of the Lagrangian (see Ref. \cite{hh})
\begin{align}
L=L_g+L_m,
\end{align}
where
\begin{align}
L_g = \frac1{16\pi G}\int\!d^3\!x\,N\sqrt h\left[K_{ij}K^{ij}-\left({K^i}_{i}\right)^2+{}^{(3)}\!R\right],
\end{align}
and 
\begin{align}
L_m &= \frac12\int\!d^3x \frac{\sqrt h }{N} \bigg[ \bigg( \frac{d\Phi}{dt} \bigg)^{2}- 2 N^i \partial_i\Phi \frac{d\Phi}{dt} - (N^2 h^{ij}- N^iN^j )\partial_i\Phi \partial_j\Phi - N^2 m^2 \Phi^2 \bigg].
\end{align}
For convenience, we adopt the Arnowitt-Deser-Misner (ADM) decomposition for our model with $h_{ij}$ the three-metric, $N$ the lapse function and $N^i$ the shift vector, all of them at a given time $t$. Here, the spatial indices $i,j, \ldots$ are denoted with Latin letters and they run from 1 to 3.
In addition, $\sqrt{h}$ is the determinant of the spatial metric, $K_{ij}$ the corresponding extrinsic curvature and ${}^{(3)}\!R$ the Ricci scalar of the spatial sections. The homogeneous counterpart of the spacetime metric is determined by $\alpha(t)$, the logarithm of the scale factor $a(t)$, and the homogeneous lapse $N_0(t)$. 
We only introduce small tensor perturbations. For this isotropic spacetime (and within the truncation scheme we will adopt for them in this manuscript) they decouple from scalar and vector modes. In what follows, we will focus the analysis on tensor perturbations, ignoring scalar and vector ones. In order to describe the inhomogeneities, it will be convenient to introduce the real tensor harmonics $\tilde G_{ij}$ (see Appendix \ref{app:ten-harm}). They are transverse $^0h^{ij}\;^{0}\nabla_i \tilde G_{jk}\,=0$ and traceless $^0h^{ij}\tilde G_{ij}=0 $ and satisfy
\begin{equation}
^0h^{ij}\;^{0}\nabla_i^{0}\nabla_j \tilde G_{kl}= -\omega_n^2\tilde G_{kl},
\end{equation}
where $^{0}\nabla_i$ is the covariant derivative with respect to a static auxiliary three-metric $^0h_{ij}$  that, in the present compact flat Universe, can be chosen as the standard flat metric on the three-torus. Each orthonormal direction has coordinate $\theta_i$ such that $2\pi \theta_i/ l_0 \in S^1$ with period equal to $l_0$. Besides, $\omega_n^2=4\pi^2\vec n\cdot\vec n/l_0^{2}$ where $\vec n=(n_1,n_2,n_3)\in\mathbb Z^3$ is any tuple of 
integers. Since there is no zero mode corresponding to tensor perturbations, the tuple $\vec n =  0$ will be excluded in the expansion. Let us notice that, since these modes are transverse and traceless, there are only two linearly independent tensor modes (one per polarization). Furthermore, each real harmonic $\tilde G_{kl}$ is odd or even under the symmetry transformation $\theta_i\to(l_0-\theta_i)$. Hence, they must be labeled with the tuple $\vec n$ such that the first nonvanishing component is strictly positive, in order to avoid repetitions. In total, we have $(\tilde G_{ij})_{\vec {\bf n}}$, where the subscript $\vec {\bf n}=(\vec n,\epsilon,\tilde{\epsilon})$ codifies the tuple $\vec n$, the parity $\epsilon=\pm$ and the polarization $\tilde{\epsilon}=(+,\times)$.

This set of tensor harmonics provides a basis to expand pure tensor functions. In particular, if we ignore scalar and vector perturbations, the metric takes the form 
\begin{align}\label{eqs:expansions}
&h_{ij}(t,\vec\theta) = \sigma^2 e^{2\alpha(t)}\left(\;{}^0h_{ij}(\vec\theta)+2\sqrt{6}\sum_{\vec {\bf n}}d_{\vec {\bf n}} (t)(\tilde G_{ij})_{\vec {\bf n}}(\theta)\right).
\end{align}
Here, $\sigma^2=4\pi G/(3l_0^3)$, and $G$ is the Newton constant. 

The scalar field, within our truncation scheme for the perturbations, is written as
\begin{equation}
\Phi(t,\vec\theta) = \frac{1}{\sigma\sqrt{l_0^{3}}}\varphi(t).
\end{equation}

We now substitute these expressions in the action and truncate the result at quadratic order in the inhomogeneities. We obtain a total Hamiltonian $H$ which is constrained to vanish, i.e.
\begin{equation}\label{eq:Hamiltonian}
H = N_0\Big[H_{|0}+\sum_{\vec {\bf n}}\;{^{T}\!H^{\vec {\bf n}}_{|2}}\Big],
\end{equation}
that corresponds to the homogeneous mode of the scalar constraint (within the quadratic truncation in the perturbations and neglecting scalar and vector ones)  of the full theory (see Appendix B of Ref. \cite{hh}). Here, ${^{T}\!H^{\vec {\bf n}}_{|2}}$ is quadratic in the tensor perturbations. No further linear constraints appear. This is due to the fact that there is no coupling with tensor matter fields. In this case, we are left in the inhomogeneous sector with two local degrees of freedom. 

More specifically, 
\begin{equation}\label{eq:H_0}
H_{|0} = \frac{e^{-3\alpha}}{2}\big(\pi_\varphi^2-{\cal H}_0^{(2)}\big),
\end{equation}
with
\begin{align}\label{eq:H0_2}
	{\cal H}_0^{(2)}= \pi_{\tilde \alpha}^2-e^{6\tilde{\alpha}}\bar m^2\varphi^2,
\end{align}
where the constant $\bar m$ is related to the mass $m$ of the scalar field by $\bar m= m \sigma$. In addition, $\pi_\alpha$ and $\pi_\varphi$ are the momenta conjugate to the variables $\alpha$ and $\varphi$, respectively. Furthermore, 
\begin{align}\label{eq:TH_2}
{^{T}\!H^{\vec {\bf n}}_{|2}} &= \frac{1}{2}e^{-3\alpha}\big[\pi_{d_{\vec {\bf n}}}^2+8\pi_\alpha d_{\vec {\bf n}}\pi_{d_{\vec {\bf n}}}+2\left(5{\cal H}_0^{(2)}+3\pi_\varphi^2+2e^{6\alpha}\bar m^2\varphi^2\right)d_{\vec {\bf n}}^2+e^{4\alpha}\omega_n^2d_{\vec {\bf n}}^2\big],
\end{align}
with $\pi_{d_{\vec {\bf n}}}$ the conjugate variable of $d_{\vec {\bf n}}$ (see Refs. \cite{hh,mikel} for further details). Let us notice that these mode functions are gauge invariant in the sense of Bardeen's potentials \cite{bardeen}. 

We will now introduce a canonical transformation in the system that will simplify the classical analysis but also will have important consequences in the quantization that we will consider below. It is given by a redefinition of the tensor perturbations by means of a scaling with the scale factor (typically adopted in these systems) and a redefinition of its momentum as
\begin{align}\label{eq:can-trans}
\tilde d_{\vec {\bf n}}=e^{\alpha}  d_{\vec {\bf n}},\quad \tilde\pi_{\tilde d_{\vec {\bf n}}}=e^{- \alpha}\big(\pi_{ d_{\vec {\bf n}}}+3\pi_{\tilde \alpha} d_{\vec {\bf n}}\big).
\end{align}
This canonical transformation must be extended to the homogeneous sector by means of backreaction contributions as
\begin{align}
\tilde \alpha=\alpha-\frac{3}{2}\sum_{\vec {\bf n}} d_{\vec {\bf n}}^2,\quad \pi_{{\tilde  \alpha}}=\pi_{\alpha}-\sum_{\vec {\bf n}} d_{\vec {\bf n}}\big(\pi_{ d_{\vec {\bf n}}}+3\pi_{\tilde \alpha} d_{\vec {\bf n}}\big).
\end{align}
We must notice that the new variables are still homogeneous degrees of freedom. The scalar field $\varphi$ and its momentum $\pi_\varphi$ are unaltered by this transformation. Besides, it is easily invertible keeping in mind the truncation scheme we adopt here. This new set of modes is usually known as Mukhanov-Sasaki tensor variables. The old variables can be written in terms of these new ones (neglecting high-order perturbation contributions) as
\begin{subequations}
\begin{align}\label{eq:can-trans1}
&d_{\vec {\bf n}}=e^{-\tilde \alpha} \tilde d_{\vec {\bf n}},\quad \pi_{d_{\vec {\bf n}}}=e^{\tilde \alpha}\big(\pi_{\tilde  d_{\vec {\bf n}}}-3\pi_{\tilde \alpha} e^{-2\tilde \alpha} \tilde d_{\vec {\bf n}}\big),\\
&\alpha=\tilde \alpha+\frac{3}{2}e^{-2\tilde \alpha}\sum_{\vec {\bf n}}\tilde d_{\vec {\bf n}}^2,\quad \pi_{{\alpha}}=\pi_{\tilde \alpha}+\sum_{\vec {\bf n}}\tilde d_{\vec {\bf n}}\pi_{\tilde  d_{\vec {\bf n}}}.
\end{align}
\end{subequations}

The unperturbed Hamiltonian $H_{|0}$ has the same functional form as Eq. \eqref{eq:H_0} in terms of the new variables $(\tilde \alpha, \pi_{\tilde \alpha})$, but the quadratic contribution in the perturbations to the scalar constraint now becomes 
\begin{equation}\label{eq:barTH_2}
{^{T}\!{\bar H}^{\vec {\bf n}}_{|2}} = {^{T}\!{\tilde H}^{\vec {\bf n}}_{|2}}+\frac{3}{2}e^{-2\tilde \alpha}\tilde  d_{\vec {\bf n}}^2 H_{|0}, 
\end{equation}
with
\begin{align}\label{eq:tildeTH_2}
{^{T}\!{\tilde H}^{\vec {\bf n}}_{|2}} &= \frac{1}{2}e^{-\tilde \alpha}\left[\pi_{\tilde d_{\vec {\bf n}}}^2+\left(e^{-4\tilde \alpha}{\cal H}_0^{(2)}-2e^{2\tilde\alpha}\bar m^2\varphi^2+\omega_n^2\right){\tilde d}_{\vec {\bf n}}^2\right].
\end{align}

Finally, the second term in Eq. \eqref{eq:barTH_2} can be absorbed in a redefinition of the homogeneous lapse function as 
\begin{equation}
\tilde N_0 = \left(1+\frac{3}{2}e^{-2\tilde \alpha}\sum_{\vec {\bf n}}{\tilde d}_{\vec {\bf n}}^2\right)N_0.
\end{equation}

The final form of the Hamiltonian, truncating the perturbations to second order in the action, is given by $ H=\tilde N_0\Big[H_{|0}+\sum_{\vec {\bf n}}\;{^{T}\!{\tilde H}^{\vec {\bf n}}_{|2}}\Big]$. 

\section{Quantization}\label{sec:quantization}

In this section, we adopt the hybrid quantization approach of Ref. \cite{muk-hyb} for our model. Given a quantum theory of gravity, the nature of the hybrid quantization resides in the existence of an intermediate regime between full quantum gravity and the semiclassical approximation of a QFT in curved spacetimes\footnote{In this case, the spacetime is described by a set of classical trajectories, or semiclassical ones that in principle can incorporate corrections coming from quantum gravity.}. In this situation, we expect that the main contribution of quantum gravity will be codified in the background variables, while the inhomogeneities, though they codify geometrical degrees of freedom, can be treated as standard quantum fields. We will then consider a generic quantization of our homogeneous sector, i.e. for the geometrical variables $(\tilde \alpha,\pi_{\tilde \alpha})$ and the matter field $(\varphi,\pi_\varphi)$, where the respective kinematical Hilbert spaces are ${\cal H}_{\rm kin}^{\rm grav}$ and ${\cal H}_{\rm kin}^{\rm matt}$. In Sec. \ref{app:polym}, we consider a concrete (polymer) representation for the background geometry. Regarding the inhomogeneities, we will adopt a standard Fock quantization with the kinematical Hilbert space denoted by ${\cal F}$. The particular representation will not be specified yet. The kinematical Hilbert space of the system ${\cal H}_{\rm kin}$ will be the tensor product ${\cal H}_{\rm kin}= {\cal H}_{\rm kin}^{\rm grav}\otimes{\cal H}_{\rm kin}^{\rm matt}\otimes{\cal F}$. For simplicity, we fix the reduced Planck constant $\hbar$ equal to the unit in all our discussion.

Let us start by reviewing the quantization of the homogeneous sector. Our analysis (as well as the one in Refs. \cite{AAN2,AM,muk-hyb}) will be based in the choice of $\varphi$ as physical time in the unperturbed quantum theory\footnote{In the present model, this choice of time is expected to be admissible only for some intervals of the evolution since $\varphi$ is not monotonous on shell. However, close to the high quantum region (at the bounce), it will be a good choice for states where the energy density is kinematically dominated. Other choices of time have been explored in the literature \cite{B-phys-time}.}. This evolution picture in cosmology is commonly used since the constraint equation for the physical states $\Psi^0$ of the exact homogeneous model resembles a Klein-Gordon equation with $\varphi$ playing the role of time. In this case, one chooses $\varphi$ as a time parameter, and its conjugate momentum is $\hat\pi_{\varphi}=-i\partial_\varphi$. More precisely,
\begin{equation}
\hat\pi_{\varphi}^2\Psi^0-\hat{\cal H}_0^{(2)}\Psi^0=0,
\end{equation}
with $\hat{\cal H}_0^{(2)}=\hat \pi_{\tilde \alpha}^2-\widehat{e^{6\tilde{\alpha}}}\bar m^2\hat\varphi^2$. It is worth commenting that, as we will see below, in the case in which the scalar field is massless the evolution is well defined globally since the square root of $\hat{\cal H}_0^{(2)}$ would be time independent. This is not the case for the present massive scalar field. 

Let us assume that there exists a well-defined unitary evolution operator with respect to this choice of time, at least locally. The states at $\varphi_0$ and $\varphi$ are related such that
\begin{equation}\label{eq:hom-sol}
\Psi^0_\varphi = \hat U(\varphi,\varphi_0) \Psi^0_{\varphi_0},
\end{equation}
where $\hat U(\varphi,\varphi_0)$ is defined in the standard way in quantum mechanics by means of the condition $[\hat \pi_\varphi,\hat U]={\hat{{\cal H}}_0}\hat U$, where ${\hat{{\cal H}}_0}$ is self-adjoint. The operator $\hat U(\varphi,\varphi_0)$ can be then written as 
\begin{align}
\hat U(\varphi,\varphi_0)={\cal P}\left[\exp\left(\int_{\varphi_0}^\varphi \,d\tilde\varphi {\hat{{\cal H}}_0}(\tilde \varphi)\right)\right],
\end{align}
with ${\cal P}$ the usual time ordering with respect to $\varphi$. One can easily see that the operators ${\hat{{\cal H}}_0}$ and $\hat{\cal H}_0^{(2)}$ fulfill
\begin{equation} \label{eq:hom-consist}
\hat{\cal H}_0^{(2)}=({\hat{{\cal H}}_0})^2+[\hat \pi_\varphi,{\hat{{\cal H}}_0}].
\end{equation}
Let us briefly comment that, as we already anticipated, if $\hat{\cal H}_0^{(2)}$ is time independent, the previous equation admits solutions of the form $[\hat \pi_\varphi,{\hat{{\cal H}}_0}]=0$, and the evolution operator is determined by the self-adjoint true Hamiltonian ${\hat{{\cal H}}_0}=\sqrt{|\hat{\cal H}_0^{(2)}|}$. For instance, this is the situation for a massless scalar field. But in the present model, we still lack an exact expression for ${\hat{{\cal H}}_0}$. 

Moreover, when the inhomogeneities are incorporated, the analysis becomes even more intriguing. Following the notation in Ref. \cite{muk-hyb}, let us define the classical phase space functions
\begin{subequations}
\begin{align}
{\vartheta} &= e^{2\tilde \alpha},\\
{\vartheta}_T^q &= e^{-2\tilde \alpha}{\cal H}_0^{(2)}-2e^{4\tilde{\alpha}}\bar m^2\varphi^2,
\end{align}
\end{subequations}
together with 
\begin{equation}
{\Theta}^T = -\sum_{\vec {\bf n}}\left[\left({\vartheta}\omega_n^2+{\vartheta}_T^q \right){\tilde d}_{\vec {\bf n}}^2 + {\vartheta} \pi_{\tilde d_{\vec {\bf n}}}^2\right].
\end{equation}
This last quantity is proportional to the (densitized) homogeneous contribution of the tensor modes to the scalar constraint (within our truncation approximation), i.e.
\begin{equation}
2 e^{3\tilde \alpha}\sum_{\vec {\bf n}}{^{T}\!{\tilde H}^{\vec {\bf n}}_{|2}} = - {\Theta}^T.
\end{equation}

Following the hybrid approach, in order to represent the inhomogeneities, we adopt a Fock quantization. Among the different representations that one can choose, we select the ones that, within the semiclassical approximation of QFT in curved spacetimes, satisfy i) the spatial isometries leave invariant the vacuum state and ii) the semiclassical dynamics of the tensor modes being unitarily implementable. These criteria leave us with a unique class of unitarily equivalent Fock quantizations  \cite{uniqueness1,uniqueness2}. It is also important to notice that the canonical pair of variables that we have chosen for the description of the modes of the perturbations is the only one (up to unitary equivalence) compatible with those representations \cite{uniqueness1}. In other words, if we would have not implemented the transformation given in Eq. \eqref{eq:can-trans}, we would have not been able to find a Fock quantization in terms of $d_{\vec {\bf n}}$ and $\pi_{d_{\vec {\bf n}}}$ with unitary dynamics. In particular, within this family we can choose the representation where the annihilationlike operators
\begin{equation}\label{annihi}
\hat a_{\tilde d_{\vec {\bf n}}}=\frac{1}{\sqrt{2 \omega_n}}(\omega_n \hat{\tilde d}_{\vec {\bf n}}+i\hat \pi_{\tilde d_{\vec {\bf n}}}),
\end{equation}
and their adjoint $\hat a^\dagger_{\tilde d_{\vec {\bf n}}}$ (creationlike operators), are those naturally associated with harmonic oscillators of frequency $\omega_n$. It is usually called the massless representation. All these questions about the Fock quantization of tensor modes were discussed in Ref. \cite{mikel}.

With this in mind, and adopting the previous massless representation for the tensor modes, we can define the operator corresponding to the densitized scalar constraint as
\begin{equation}\label{eq:q-cont}
\hat{\tilde H}=\frac{1}{2}\left(\hat\pi_{\varphi}^2-\hat{\cal H}_0^{(2)}-\hat{\Theta}^T\right).
\end{equation}
We assume that the quantum operators involved in this expression can be densely defined on ${\cal H}_{\rm kin}$. 

\section{Born-Oppenheimer ansatz}\label{BOa}

The previous quantum operator represents the homogeneous mode of the scalar constraint of our system once the action is truncated to second order in the perturbations. Besides, the constraint algebra is free of anomalies (the quantum constraint commutes with itself). Therefore, one can adopt the Dirac quantization approach in order to analyze the quantum dynamics of the system. It involves looking for the kernel of the operator $\hat{\tilde H}$. According to it, one can easily realize that the coupling between the homogeneous and inhomogeneous sectors given by $\hat{\Theta}^T$ is highly nontrivial. Therefore, the solutions will not keep the structure of the tensor product form of the kinematical Hilbert space ${\cal H}_{\rm kin}$.  It is remarkable that in LQC the physical states can be determined out of suitable initial data and they can be endowed with an inner product, and observables acting on this space can be formally constructed (see Ref. \cite{hybrid}). The current numerical and analytical tools must be developed in order to construct exact solutions to the constraint. Instead, since we are mainly interested in the semiclassical sector of the theory, we will consider here analytical approximations, valid for any representation. For instance, if the homogeneous and inhomogeneous sectors present a different rate of variation with respect to the physical clock $\varphi$, we can adopt a Born-Oppenheimer ansatz for the solutions of the form
\begin{equation}\label{BOans}
\Psi_\varphi(\alpha,\tilde d_{\vec {\bf n}})=\Gamma_\varphi(\alpha)\psi_\varphi(\tilde d_{\vec {\bf n}}),
\end{equation}
where $\Gamma_\varphi$ is a normalized state defined on the homogeneous sector ${\cal H}_{\rm kin}^{\rm grav}$ while $\psi_\varphi$ belongs to ${\cal F}$. In this case, the operator $\hat \varphi$ (and its different powers) is again regarded as a time parameter, and $\hat \pi_\varphi$ is regarded as $(-i)$ times the derivative with respect to it $\varphi$.

In the presence of inhomogeneities, and under this Born-Oppenheimer ansatz, the time-dependent states $\Gamma_\varphi$ are such that Eq. \eqref{eq:hom-consist} might not be fulfilled. Indeed, we only require that the evolution is dictated by a self-adjoint operator $\hat{\tilde{{\cal H}}}_0$. They will be considered an approximate solution to the homogeneous sector whenever $({\hat{\tilde{{\cal H}}}_0})^2-\hat{\cal H}_0^{(2)}+[\hat \pi_\varphi,{\hat{\tilde{{\cal H}}}_0}]$ is negligibly small with respect to some other quantities, as we will see below. 

If we apply the quantum constraint \eqref{eq:q-cont} to this ansatz, with $\Gamma_\varphi$ fulfilling Eq. \eqref{eq:hom-sol}, and we impose it to vanish, one can easily see that
\begin{align}\label{eq:BOans-const}
&\left\{\left(({\hat{\tilde{{\cal H}}}_0})^2-\hat{\cal H}_0^{(2)}+[\hat \pi_\varphi,{\hat{\tilde{{\cal H}}}_0}]\right)\Gamma_\varphi\right\}\psi_\varphi+2({\hat{\tilde{{\cal H}}}_0}\Gamma_\varphi)(\hat \pi_\varphi\psi_\varphi)+\Gamma_\varphi(\hat \pi_\varphi^2\psi_\varphi)-\hat{\Theta}^T (\Gamma_\varphi\psi_\varphi)=0,
\end{align}
taking into account that
\begin{subequations}
\begin{align}
\hat{\pi}_{\varphi}\Psi_\varphi=&\Gamma_\varphi (\hat{\pi}_{\varphi}\psi_\varphi)+(\hat{\tilde{\mathcal H}}_0\Gamma_\varphi)\psi_\varphi,\\
\hat{\pi}_{\varphi}^2\Psi_\varphi=&\Gamma_\varphi (\hat{\pi}_{\varphi}^2\psi_\varphi)+2 (\hat{\tilde{\mathcal H}}_0\Gamma_\varphi) (\hat{\pi}_{\varphi}\psi_\varphi)+([\hat{\pi}_{\varphi},\hat{\tilde{\mathcal H}}_0]\Gamma_\varphi)\psi_\varphi+ \big\{(\hat{\tilde{\mathcal H}}_0)^2\Gamma_\varphi\big\}\psi_\varphi.
\end{align} 
\end{subequations}
Here, we assume that the commutator $[\hat \pi_\varphi,{\hat{\tilde{{\cal H}}}_0}]$ is independent of $\hat \pi_\varphi$ but may depend on time, i.e. on $\varphi$. 

If we now compute the inner product on the left-hand side of \eqref{eq:BOans-const} with the bra associated to $\Gamma_\varphi$, we obtain 
\begin{align}\label{eq:BOans-constGamma}
&\hat \pi_\varphi^2\psi+2\langle{\hat{\tilde{{\cal H}}}_0}\rangle_{\Gamma}\hat \pi_\varphi\psi=\left\{\langle\hat{\cal H}_0^{(2)}-({\hat{\tilde{{\cal H}}}_0})^2+i{\rm d_\varphi}{\hat{\tilde{{\cal H}}}_0}\rangle_{\Gamma}\right\}\psi+\langle\hat{\Theta}^T\rangle_{\Gamma} \psi.
\end{align}
Here, we have introduced the definition
\begin{equation}
-i{\rm d}_\varphi\hat O=[\hat{\pi_\varphi}-{\hat{\tilde{{\cal H}}}_0},\hat O]
\end{equation}
for any operator $\hat O$, and we have dropped the subindex $\varphi$ of $\Gamma_\varphi$ in the expectation values  $\langle\hat O\rangle_\Gamma$ in order to simplify the notation. But we must keep in mind that the expectation values $\langle\hat O\rangle_\Gamma$ are time dependent. We will drop it as well in the following. 

Equation \eqref{eq:BOans-constGamma} will codify the relevant information about the dynamics of the system (under this Born-Oppenheimer ansatz) in the situation in which quantum transitions from $\Gamma_\varphi$ to other states can be neglected. This condition is fulfilled, for instance, if the operators ${\hat{\tilde{{\cal H}}}_0}$, $\hat\vartheta$, $\hat\vartheta_T^q$ and $(\hat{\cal H}_0^{(2)}-({\hat{\tilde{{\cal H}}}_0})^2-i{\rm d_\varphi}{\hat{\tilde{{\cal H}}}_0})$ have small relative dispersions on the states $\Gamma_\varphi$ for all relevant values of $\varphi$. However, we must keep in mind that there could be states where these relative dispersions are not small but the mentioned transitions are still negligible.

\subsection{Dressed metric formalism}

The dressed metric approach of Refs. \cite{AAN1,AAN2,AM} emerges in this framework if  additional conditions are met. Let us recall that the dressed metric approach is determined by a first-order Schr\"odinger evolution equation, rather than second order, as in Eq. \eqref{eq:BOans-constGamma}. In this situation, we can interpret one of the counterparts of this equation as the Hamiltonian generating the evolution of the perturbations (although we do not prove it is self-adjoint). We will analyze the additional required approximations in order to achieve this regime within the present formalism. The first step in our analysis is to find the conditions that allow us to neglect the contribution $\hat{\pi}_{\tilde\varphi}^2 \psi$ in Eq. \eqref{eq:BOans-constGamma}. Let us start by acting with $\hat\pi_\varphi$ on Eq. \eqref{eq:BOans-constGamma}, which yields
\begin{align}\nonumber\label{eq:deriv}
&\left[\frac{3i\langle {\mathrm d}_{\tilde\varphi}\hat{\tilde{\mathcal H}}_0 \rangle_{\Gamma}+\langle \hat{\mathcal H}_0^{(2)} - (\hat{\tilde{\mathcal H}}_0)^2 \rangle_\Gamma}{2\langle \hat{\tilde{\mathcal H}}_0 \rangle_{\Gamma}} +2  \langle \hat{\tilde{\mathcal H}}_0 \rangle_{\Gamma}\right] \hat{\pi}_{\tilde\varphi}^2\psi 
= \frac{2i\langle {\mathrm d}_{\tilde\varphi}\hat{\tilde{\mathcal H}}_0 \rangle_{\Gamma}+\langle \hat{\mathcal H}_0^{(2)} - (\hat{\tilde{\mathcal H}}_0)^2 \rangle_\Gamma}{2\langle \hat{\tilde{\mathcal H}}_0 \rangle_{\Gamma}} \nonumber\\
&\times\left[2\langle \hat{\Theta}^T\rangle_{\Gamma}
+\frac{i}{2} \langle   3{\mathrm d}_{\tilde\varphi}\hat{\tilde{\mathcal H}}_0 \rangle_{\Gamma}+\frac5{4}\langle \hat{\mathcal H}_0^{(2)} - (\hat{\tilde{\mathcal H}}_0)^2 \rangle_\Gamma\right] \psi 
-\frac1{8}\frac{(\langle \hat{\mathcal H}_0^{(2)} - (\hat{\tilde{\mathcal H}}_0)^2 \rangle_\Gamma)^2}{\langle \hat{\tilde{\mathcal H}}_0 \rangle_{\Gamma}}\psi- \hat{\pi}_{\tilde\varphi}^3 \psi
\nonumber\\
& 
-i
\left[\langle {\mathrm d}_{\tilde\varphi}\hat{\Theta}^T\rangle_{\Gamma}
+i \langle   {\mathrm d}_{\tilde\varphi}^2\hat{\tilde{\mathcal H}}_0 \rangle_{\Gamma}+\langle {\mathrm d}_{\tilde\varphi} \hat{\mathcal H}_0^{(2)} - {\mathrm d}_{\tilde\varphi}(\hat{\tilde{\mathcal H}}_0)^2 \rangle_\Gamma\right] \psi.
\end{align}
Then, the contribution $\hat{\pi}_{\tilde\varphi}^2 \psi$ in Eq. \eqref{eq:BOans-constGamma} is negligible if 
\begin{itemize}
 \item[i)] $\langle \hat{\mathcal H}_0^{(2)} - (\hat{\tilde{\mathcal H}}_0)^2 \rangle_\Gamma$ is negligible with respect to linear terms in the perturbations,
  \item[ii)] $\langle {\mathrm d}_{\tilde\varphi} \hat{\mathcal H}_0^{(2)} - {\mathrm d}_{\tilde\varphi}(\hat{\tilde{\mathcal H}}_0)^2 \rangle_\Gamma$, $\langle {\mathrm d}_{\tilde\varphi}\hat{\Theta}^T\rangle_{\Gamma}$ and $\langle {\mathrm d}^2_{\tilde\varphi}\hat{\tilde{\mathcal H}}_0 \rangle_{\Gamma}$ are negligible compared to terms of quadratic order in the perturbations,
 \item[iii)] $\langle {\mathrm d}_{\tilde\varphi}\hat{\tilde{\mathcal H}}_0 \rangle_{\Gamma}$ is at most of the order of quadratic terms.
\end{itemize}
In addition, we will assume that we can neglect $\hat{\pi}_{\tilde\varphi}^3 \psi$ in Eq. \eqref{eq:deriv}. To check whether this is possible, we only need to operate repeatedly with  $\hat{\pi}_{\tilde\varphi}$ on $\psi$ and check if $\hat{\pi}_{\tilde\varphi}^{n+1}$ is negligible compared to the action of $\hat{\pi}_{\tilde\varphi}^{n}$. This happens in particular for $n=1$ if we can neglect $\hat{\pi}_{\tilde\varphi}^3 \psi$ in Eq. \eqref{eq:deriv}, together with the previous conditions i to iii.

Then, dropping $\hat{\pi}_{\tilde\varphi}^2 \psi$ in Eq. \eqref{eq:BOans-constGamma} and dividing by $2\langle{\hat{\tilde{{\cal H}}}_0}\rangle_\Gamma$, we are left with
\begin{align}\label{eq:BOans-constGamma1}
&\hat \pi_\varphi\psi=\frac{\langle\hat{\Theta}^T\rangle_\Gamma}{2\langle{\hat{\tilde{{\cal H}}}_0}\rangle_\Gamma} \psi+\frac{1}{2\langle{\hat{\tilde{{\cal H}}}_0}\rangle_\Gamma}
\left\{\langle\hat{\cal H}_0^{(2)}-({\hat{\tilde{{\cal H}}}_0})^2+i{\rm d_\varphi}{\hat{\tilde{{\cal H}}}_0}\rangle_\Gamma\right\}\psi.
\end{align}
Now, we can introduce several additional approximations. One possibility is that $\langle i{\rm d_\varphi}{\hat{\tilde{{\cal H}}}_0}\rangle_\Gamma$ is negligible compared with contributions quadratic in the perturbations. Otherwise, we will have a contribution in our Schr\"odinger equation potentially spoiling the unitarity of the evolution. In this case, we are left with 
\begin{align}\label{eq:BOans-constGamma2}
&\hat \pi_\varphi\psi=\frac{\langle\hat{\Theta}^T\rangle_\Gamma}{2\langle{\hat{\tilde{{\cal H}}}_0}\rangle_\Gamma} \psi+\frac{1}{2\langle{\hat{\tilde{{\cal H}}}_0}\rangle_\Gamma}
\left\{\langle\hat{\cal H}_0^{(2)}-({\hat{\tilde{{\cal H}}}_0})^2\rangle_\Gamma\right\}\psi.
\end{align}
An additional approximation would be to ask that the quantity $\langle\hat{\cal H}_0^{(2)}-({\hat{\tilde{{\cal H}}}_0})^2\rangle_\Gamma$ is negligible with respect to $\langle\hat{\Theta}^T\rangle_\Gamma$. Then, the previous expression reduces to 
\begin{align}\label{eq:BOans-constGamma3}
\hat \pi_\varphi\psi=\frac{\langle\hat{\Theta}^T\rangle_\Gamma}{2\langle{\hat{\tilde{{\cal H}}}_0}\rangle_\Gamma} \psi,
\end{align}
which is the analogous expression that appears in the dressed metric approach.  

All these approximations are in agreement with the ones assumed in Ref. \cite{muk-hyb} for scalar perturbations. Our analysis shows for the first time the explicit relation of the hybrid quantization with the dressed metric approach of Refs. \cite{AAN1,AAN2,AM} for tensor perturbations.

\subsection{Effective equations for the tensor perturbations}\label{MSequations}

We will now provide the effective equations of motion for tensor perturbations within the hybrid approach. The dynamics of the system, within the Born-Oppenheimer ansatz and neglecting transitions of background states by means of the scalar constraint, is governed by the quantum Hamiltonian constraint in Eq. \eqref{eq:BOans-constGamma}. All the operators there are defined in the kinematical Hilbert space ${\cal F}$, where we must recall that $\varphi$ is the time parameter. In order to provide the effective equations of motion, we will naively replace all these operators by their classical analogs. We then obtain an effective constraint $({\cal C}_{T,\Gamma}^{\rm eff})/2$ generating the effective evolution of the perturbations ${\tilde d_{\vec {\bf n}}} $ and $\pi_{\tilde d_{\vec {\bf n}}} $. Let us notice that it is naturally adapted to the effective harmonic time $\bar T_\Gamma$ that is related with the effective cosmic time $t_\Gamma$ by means of $dt_\Gamma=\sigma \langle e^{3\hat\alpha}\rangle_\Gamma d\bar T_\Gamma$ and with the conformal time through $l_0 d\eta_\Gamma=\langle \hat\vartheta\rangle_\Gamma d\bar T_\Gamma$. These relations can be easily deduced just by noticing that the Hamiltonian constraint in Eq. \eqref{eq:q-cont} corresponds to a scaled version of the original one by a factor $e^{3\alpha}$. In conformal time, for instance, we have 
\begin{align}\label{eq:can-eff}
\frac{d}{ d\eta_\Gamma}{\tilde d}_{\vec {\bf n}}=l_0\pi_{{\tilde d}_{\vec {\bf n}}}, \quad \frac{d}{ d\eta_\Gamma}\pi_{{\tilde d}_{\vec {\bf n}}}= -l_0{\tilde d}_{\vec {\bf n}}\left(\omega_n^2+\frac{\langle\hat{\vartheta}_T^q\rangle_\Gamma}{\langle\hat{\vartheta}\rangle_\Gamma}\right).
\end{align}
Combining these two equations, we get
\begin{equation}\label{eq:eff-eom}
\frac{d^2}{ d\eta_\Gamma^2}{\tilde d}_{\vec {\bf n}}= -{\tilde d}_{\vec {\bf n}}\left(\tilde\omega_n^2+\langle\hat\theta^T_\Gamma\rangle_\Gamma\right),
\end{equation}
such that $\tilde\omega_n^2=l_0^2\omega_n^2$ and
\begin{equation}\label{eq:exp-theta}
\langle\hat\theta^T_\Gamma\rangle_\Gamma=l_0^2\frac{\langle\hat{\vartheta}_T^q\rangle_\Gamma}{\langle\hat{\vartheta}\rangle_\Gamma}.
\end{equation}

These effective equations of motion are hyperbolic in the ultraviolet regime since $\tilde\omega_n^2$ would be the dominant term inside the parenthesis in Eq. \eqref{eq:eff-eom}. Besides, the other contribution, i.e. $\langle\hat\theta^T_\Gamma\rangle_\Gamma$, incorporates quantum geometry corrections that can be crucial at the Planck regime, where classical general relativity is expected to break down. It is important to keep in mind that the definition of time (either a harmonic, cosmic or conformal one) is intrinsically linked to the background state $\Gamma_\varphi$ in this effective description. Similar relations were also found in the case of scalar perturbations \cite{muk-hyb}. Notice that the operator $\hat{\vartheta}$ is the same for both scalar and tensor perturbations, and so is its expectation value on the state $\Gamma_\varphi$. Hence, the definitions of harmonic, cosmic and conformal times are common to both types of perturbations. In addition, let us recall that within this Born-Oppenheimer ansatz the only required approximations are that the transitions of background states by means of the scalar constraint are negligible and that we can replace the quantum operators in Eq. \eqref{eq:BOans-constGamma} by their classical expressions.

Within the dressed metric regime, although the effective equations of motion have an explicit form that coincides with the one given in Eq. \eqref{eq:eff-eom}, due to the approximations required to obtain Eq. \eqref{eq:BOans-constGamma3} from Eq. \eqref{eq:BOans-constGamma}, the specific expectation values of Eq. \eqref{eq:exp-theta} as well as the times $t_\Gamma$, $T_\Gamma$ and $\eta_\Gamma$ will be affected. One of the advantages of this dressed metric formalism is that one has a true Hamiltonian generating the evolution of the tensor modes. In this situation, if we work in the Heisenberg picture, the expectation values of the operators $\hat{\tilde d}_{\vec {\bf n}} $ and $\hat \pi_{\tilde d_{\vec {\bf n}}} $ on an arbitrary state $\psi$ fulfill the classical equations of motion. This is guaranteed by Erhenfest's theorem and the fact that the true Hamiltonian is quadratic in the perturbations (provided it is self-adjoint). Then, one could treat $\langle \hat{\tilde d}_{\vec {\bf n}} \rangle_{\psi} $ and $\langle\hat \pi_{\tilde d_{\vec {\bf n}}}\rangle_{\psi} $ classically. 

\subsection{LQC background quantization}\label{app:polym}

We will close this section with an explicit example where the previous results can be applied. Let us consider a quantization of the background within the framework of LQC \cite{lqc-rev}. We will adhere to the  so-called improved dynamics \cite{aps}. Concretely, in FRW cosmologies, the geometrical degrees of freedom are described by two dynamical variables, $v$ and $b$, which are canonical conjugate $\{b,v\}=4\pi G\gamma$, where $\gamma$ is the standard Immirzi parameter in LQG \cite{lqg}. The relation between these variables and the ones employed in this manuscript, i.e. $\tilde \alpha$ and its momentum $\pi_{\tilde \alpha}$, is given by
\begin{equation}\label{eq:homchange}
	|v| = l_0^3\sigma^3e^{3\tilde\alpha},\quad vb = -\gamma l_0^3\sigma^2\pi_{\tilde \alpha}.
\end{equation}
The sign of $v$ plays no role in the classical analysis. In addition, the variables used in LQC for the scalar field are proportional to $\varphi$ and $\pi_\varphi$, by means of
\begin{equation}
	\phi = \frac{\tilde \varphi}{l_0^{3/2}\sigma},\quad \pi_\phi = l_0^{3/2}\sigma\pi_{\tilde \varphi}.
\end{equation}
In order to jump to the quantization, we consider a concrete realization of the Hilbert space ${\mathcal H}_{\mathrm{kin}}^{\mathrm{grav}}$ for the homogeneous sector of the geometry. In the $v$-representation (where the operator associated to $v$ acts by multiplication), there is a basis of eigenstates $\{ |v \rangle , v\in \mathbb{R}\}$ of $\hat v$ with discrete norm $\langle v_1 | v_2 \rangle = \delta_{v_1,v_2}$. Then, we have
\begin{equation}
	\hat v |v\rangle =(2\pi G \gamma  \sqrt{\Delta})^{3/2} v |v\rangle,
\end{equation}
where $\Delta$ is the minimum nonzero eigenvalue allowed for the area operator in LQG \cite{lqg}. Together with the volume operator we also have the  displacement operator $\hat N_{\bar\mu}$, 
\begin{equation}
	\hat N_{\bar\mu}|v\rangle = |v+1\rangle.
\end{equation}
It provides the quantum representation for the matrix elements of the holonomies of the connection. They are computed along edges with length equal to $l_0\bar \mu$, with $\bar \mu=\sqrt{\Delta/p}$. They are constructed such that the physical area enclosed in a square of holonomies along edges of this kind is $\Delta$. Unlike in the full quantum theory, in homogeneous cosmologies this regulator cannot be taken to vanish. 

For the scalar field, as we did above, we adopt a standard representation such that the kinematical Hilbert space $\mathcal H_\mathrm{kin}^\mathrm{matt}$ is the space $L^2(\mathbb{R},d\phi)$ of square integrable functions on $\phi$ with the standard Lebesgue measure. In the following, we will assume the representation on which $\phi$ acts by multiplication and $\pi_{\phi}$ as $-i$ times the derivative with respect to $\phi$.

Based on the quantization prescription of Ref. \cite{MMO}, we will represent the counterpart of the zero mode of the Hamiltonian constraint corresponding exclusively to the homogeneous sector as the operator \cite{MMO,muk-hyb,hybrid,hybrid1}
\begin{equation}\label{eq:C_0}
	\hat H_{|0} = \frac{\sigma}{2}\widehat{\left[\frac1{v}\right]}^{1/2}\hat{\mathcal C}_0\widehat{\left[\frac1{v}\right]}^{1/2}.
\end{equation}
Here, we define 
\begin{align}
	\label{eq:calC_0}
	\hat{\mathcal C}_0 &= {\hat \pi}_\phi^2- \frac{4 \pi G}{3}\hat{\mathcal H}_0^{(2)}, \\	
	\hat{\mathcal H}_0^{(2)} &=  \frac{3}{4 \pi G}\left(\frac{3}{4\pi G \gamma^2}\hat\Omega_0^2-2 \hat v^2 W(\hat\phi)\right),
\end{align}
such that 
\begin{align}\label{eq:Omega}
	&\hat\Omega_0 = \frac1{4i\sqrt\Delta}\hat v^{1/2}\big[\widehat{\sgn(v)}\big(\hat N_{2\bar\mu}-\hat N_{-2\bar\mu}\big)+\big(\hat N_{2\bar\mu}-\hat N_{-2\bar\mu}\big)\widehat{\sgn(v)}\big]\hat v^{1/2}.
\end{align}
In the case of a massive scalar field, we would have $W(\hat\phi)=\frac{1}{2}m^2\hat\phi^2$. Regarding the inverse-volume operator $\widehat{[1/v]}$, it  is defined as
\begin{align}\nonumber
	&\widehat{\left[\frac1{v}\right]}^{1/3}=\frac{3}{2}\widehat{\sgn(v)}{|\hat v|}^{1/3}\big(\hat N_{-\bar\mu}{|\hat v|}^{1/3}\hat N_{\bar\mu}-\hat N_{\bar\mu}{|\hat v|}^{1/3}\hat N_{-\bar\mu}\big).
\end{align}
One can easily see that it is diagonal in the eigenbasis of the volume operator. The operator $\hat\Omega_0$ represents the classical quantity $vb$. Its contribution into the scalar constraint is by means of its square, $\hat\Omega_0^2$, which annihilates the zero-volume state $| v=0 \rangle$, leaving invariant its orthogonal complement. Since the matter sector involves the inverse-volume operator which also annihilates that state, the analysis of the solutions can be restricted to the mentioned orthogonal complement of $| v=0 \rangle$. On this complement, it is possible to restrict the study to $\hat{\cal C}_0$ thanks to the existence of a bijection between its solutions and the ones of $\hat{H}_{|0}$. In this case, the full quantization is known \cite{MMO,hybrid}. In particular, the action of  $\hat{\mathcal C}_0$ can be restricted to separable subspaces  $\mathcal H_\mathrm{\varepsilon}^\pm$ of the kinematical Hilbert space of the homogeneous geometry sector, usually called superselection sectors. The label $\varepsilon$ appears since these subspaces are formed by all the states with support on the semilattices $\mathcal L_\mathrm{\varepsilon}^\pm=\{v=\pm(\varepsilon+4n)|n\in\mathbb N\}$, where $\varepsilon\in(0,4]$. Let us remark that, on each of these subspaces, the homogeneous volume $v$ has a strictly positive minimum (or negative maximum) and its orientation does not change. We will restrict the discussion to semilattices with positive sign of $v$ in the following.

To complete the representation of the full constraint, including inhomogeneities, we have to represent the quadratic contribution of the inhomogeneities to the zero mode of the Hamiltonian constraint. For the tensor modes, we adopt again the Fock representation described above. The contributions due to the homogeneous sector will be promoted to 
\begin{align}
	\hat \vartheta&=\frac{3l_0}{4\pi G}\hat v^{2/3},\\
	\hat \vartheta_T^q&=\frac{4\pi G}{3l_0}\widehat{\left[\frac1{v}\right]}^{1/3}\hat{\mathcal H}_0^{(2)}\widehat{\left[\frac1{v}\right]}^{1/3}\!\!\!\!-\frac{4}{l_0}\hat v^{4/3}W(\hat \phi).
\end{align}

As we can see, the contribution of the tensor modes to the scalar constraint does not require additional factor orderings and polymer corrections as in the case of scalar perturbations (see Ref. \cite{muk-hyb}). The set of equations in Eq. \eqref{eq:can-eff}, or equivalently Eq. \eqref{eq:eff-eom}, is still valid in this framework. 

Although more general situations can be studied in this hybrid approach, the strategy that is commonly adopted in order to test the main quantum corrections of this formalism is to assume an effective dynamics where the corresponding quantum states are highly peaked on semiclassical trajectories, such that the expectation value of the product of basic operators can be replaced by the product of their expectation values. In this situation, if we also neglect inverse of the volume corrections, the explicit form of the effective equations of the perturbations within this effective dynamics is 
\begin{widetext}
\begin{align}\label{eq:lon-g-pert}
\frac{1}{N}\dot{\tilde d}_{\vec {\bf n}} =\frac{1}{v^{1/3}}\pi_{\tilde d_{\vec {\bf n}}},\quad \frac{1}{N}\dot{\pi}_{\tilde d_{\vec {\bf n}}} =-\frac{1}{v^{1/3}} \left(\tilde\omega_n^2-\frac{16\pi G }{3}v^{2/3}W(\phi)+\left(\frac{4 \pi G}{3}\right)^2\frac{1}{v^{4/3}}{\mathcal H}_0^{(2)}\right) {\tilde d}_{\vec {\bf n}},
\end{align}
\end{widetext}
where the dot means a derivative with respect to an arbitrary time function and $N$ the corresponding homogeneous lapse. In addition, if we neglect backreaction contributions, the effective equations of motion of the background are the usual ones \cite{as}:
\begin{subequations}\label{eq:hom-eqs2}
	\begin{align}
	\frac{1}{N}\dot{\phi} &=  \frac{\pi_\phi}{v},\\\label{eq:dot-piphi2}
	\frac{1}{N}\dot{\pi}_\phi &= -v\frac{dW(\phi)}{d\phi},
	\\\label{eq:dot-vol2}
	\frac{1}{N}\dot v &= \frac{3}{2} v\frac{\sin(2\sqrt{\Delta}\beta)}{\sqrt{\Delta}\gamma},\\
	\frac{1}{N} \dot \beta &= -\frac{3}{2} \frac{\sin^2(\sqrt{\Delta}\beta)}{\Delta\gamma}+4\pi G\gamma \left(W(\phi)-\frac{\pi_\phi^2}{2v^2}\right).
	\end{align}
\end{subequations}
In this case, we have
\begin{align}	\label{eq:H_02}
	{\mathcal H}_0^{(2)} &=  \frac{3}{4 \pi G}\left(\frac{3}{4\pi G \gamma^2}\frac{v^2\sin^2(\sqrt{\Delta}\beta)}{\Delta}-2  v^2 W(\phi)\right).
\end{align}

\subsection{Comparison with other approaches in LQC}

In order to compare the hybrid approach for cosmological perturbations with the dressed metric or the deformed algebra ones, we will show here the second-order differential equations of the Fourier modes of the tensor perturbations. 

\subsubsection{Hybrid approach}

Let us start with the hybrid approach. We combine the expressions in Eq. \eqref{eq:lon-g-pert} and replace ${\mathcal H}_0^{(2)}$ in Eq. \eqref{eq:H_02}. The result is
\begin{align}\label{eq:hyb-ten}
\tilde d_{\vec {\bf n}}'' + \left(\tilde\omega_n^2-8\pi Gv^{2/3}W(\phi)+\frac{v^{2/3}\sin^2(\sqrt{\Delta}\beta)}{\gamma^2\Delta}\right) {\tilde d}_{\vec {\bf n}}=0,
\end{align}
where the primes mean a derivation with respect to conformal time $\eta$ defined as $Ndt=v^{1/3}d\eta$. In addition, the equation of motion for the original tensor perturbation $ d_{\vec {\bf n}}$ can be straightforwardly obtained by replacing $\tilde d_{\vec {\bf n}}= v^{1/3} d_{\vec {\bf n}}$. The result is
\begin{align}\label{eq:hyb-tens}
&d_{\vec {\bf n}}'' +2{\cal H}d_{\vec {\bf n}}'+ \tilde\omega_n^2d_{\vec {\bf n}}+\left(\frac{(v^{1/3})''}{v^{1/3}}-8\pi Gv^{2/3}W(\phi)+\frac{v^{2/3}\sin^2(\sqrt{\Delta}\beta)}{\gamma^2\Delta}\right) {d}_{\vec {\bf n}}=0,
\end{align}
where we have introduced the Hubble parameter in conformal time given by
\begin{align}
{\cal H} = \frac{1}{3}\frac{v'}{v}.
\end{align}
One can see that the last time-dependent term in Eq. \eqref{eq:hyb-tens} is negligibly small away from the high curvature region, in agreement with general relativity, but it is nonvanishing close to the bounce. Indeed, in all these approaches, the effective equations of motion are in agreement with the classical theory in the low curvature limit.

\subsubsection{Dressed metric approach}

The effective equations of the tensor modes in the dressed metric approach of Ref. \cite{AAN3} can be easily read from Eqs. (2.17) and (2.18) there. They are given by
\begin{align}\label{eq:dress-ten}
d_{\vec {\bf n}}'' + 2{\cal H}d_{\vec {\bf n}}'+\tilde\omega_n^2{d}_{\vec {\bf n}}=0.
\end{align}
The form of this equation is the same than the one in general relativity. However, ${\cal H}$ incorporates quantum geometry corrections from the background at the high curvature regime. At the low curvature limit the effective dynamics of loop quantum cosmology is well approximated by classical general relativity.  

\subsubsection{Deformed algebra approach}

In the deformed algebra approach, the effective equations of motion for the tensor modes are given, for instance, in Ref. \cite{effective1}. Concretely, Eq. (34) corresponds to
\begin{align}\label{eq:def-ten}
d_{\vec {\bf n}}'' +  \left(2{\cal H}-\frac{{\bf \Omega}'}{\bf \Omega}\right)d_{\vec {\bf n}}'+{\bf \Omega}\tilde\omega_n^2{d}_{\vec {\bf n}}=0,
\end{align}
where 
\begin{align}
{\bf \Omega} = \cos(2\sqrt{\Delta}\beta)
\end{align}
is given in Eq. (29) of that reference\footnote{The background variables have been adapted to the notation of this manuscript. Besides, we consider the particular value for the parameter $k=1$ in Ref. \cite{effective1} for the sake of simplicity.}. This approach incorporates nontrivial polymeric corrections not only in ${\cal H} $, as in the previous approaches, but also through the function ${\bf \Omega}$. It is remarkable that this correction modifies the microscopic behavior of the perturbations, i.e., the ultraviolet limit of the effective equations of motion with respect to the ones in the hybrid and the dressed metric approaches. For instance, if the perturbations are evolved from a region very close to or before the bounce, they cross through a region where the effective equations of motion are not hyperbolic. Therefore, the usual prescriptions considered so far in the literature of initial data for the perturbations give power spectra at the end of inflation incompatible with observations \cite{eff-obs}. 

\subsubsection{Discussion}

The three approaches for the quantization of tensor perturbations on a loop quantum geometry show quantitative differences, which are particularly important at the deep high curvature regime. Besides, the deformed algebra approach shows important differences regarding the confrontation of predictions with observations, in comparison with the dressed metric or the hybrid approaches. At any time, these two approaches have a suitable ultraviolet behavior. On the other hand, the hybrid quantization approach incorporates quantum geometry corrections that will qualitatively affect the predictions in comparison with the dressed metric approach, at least for large scale modes, while for the ultraviolet ones we expect a good agreement, also with observations. Besides, the modification of the consistency relation $r\approx -8n_t$ between the tensor-to-scalar ratio $r$ and the tensor spectral index $n_t$ will be affected also at those scales.

\section{Conclusions}\label{conclusions}

Our purpose with this study is to complete the hybrid quantization approach applied to cosmological perturbation theory, which has been mainly focused on the study of scalar perturbations \cite{hybrid,hybrid1,muk-hyb}, since they play a prominent role in observations, while tensor modes have been always ignored. Nevertheless, the current observations provide bounds for the tensor-to-scalar ratio. The predictions of the hybrid quantization approach can be confronted with them. We have considered then a flat FRW spacetime coupled to a massive scalar field and small inhomogeneities corresponding to tensor perturbations. It is possible to carry out a separated analysis of these tensor inhomogeneities since, in this particular isotropic scenario, there is no coupling between scalar, vector and tensor perturbations in an expansion of the action to second order on them. Concretely, we assume a compact spatial topology isomorphic to a three-torus for this spacetime. We introduce small inhomogeneities by means of tensor perturbations, and we truncate the action to second order. Then, we adopt an expansion in Fourier modes for these inhomogeneities in a basis of tensor harmonics well adapted to a static flat three-metric $^0h_{ij}$. We carry out a canonical transformation in the system that not only simplifies the classical dynamics but also allows us to construct a Fock representation with unitary dynamics (in the limit of QFT in curved spacetimes). Together with the requirement of invariance under spatial symmetries, these two conditions allow us to pick out a class of unitarily equivalent Fock representations for the perturbations \cite{uniqueness1,uniqueness2}. We then adopt a quantum description of the system by means of the hybrid quantization approach \cite{hybrid,hybrid1,muk-hyb}. We study approximate solutions to the scalar constraint through a Born-Oppenheimer ansatz. We show that in this case it is sufficient to introduce some approximations in order to recover a quantum evolution equation for the perturbations of the form suggested in the dressed metric approach of Refs. \cite{AAN2,AM,AAN1}. 

In addition, we  provide the effective equations of motion of the tensor perturbations within the hybrid quantization approach. In this case, we only require within the Born-Oppenheimer ansatz that the quantum constraint produces negligible transitions between different background states and that the operators representing the tensor modes can be replaced by classical functions. In addition, we consider the particularly interesting situation  when the background geometry is represented within the framework of loop quantum cosmology \cite{lqc-rev}. Here, we restrict the study to highly peaked states. On the one hand, the effective equations of motion for the tensor perturbations incorporate the main semiclassical corrections. They will allow us to compute the power spectrum and eventually the tensor-to-scalar ratio. Its outcomes for scalar perturbations \cite{hyb-obs} allow us to conclude that it is compatible with observations \cite{precision}. On the other hand, they allow us to compare this approach with alternative ones. In particular, we have considered the dressed metric and the deformed algebra approaches in LQC. We have seen that they incorporate quantum corrections in different ways. So, we expect different quantitative predictions. For instance, the deformed algebra approach adopted in Refs. \cite{effective,effective1} seems to be ruled out \cite{eff-obs} from the point of view of the observations if one gives initial data very close to the high curvature regime or before. This is not the case of the dressed metric approach \cite{AAN1,AAN2,AM,AAN3}. A more detailed comparison among them will be a matter of future research.

\acknowledgments{We would like to thank Mercedes Mart\'in Benito and the anonymous referee for their helpful suggestions. This work was supported by Pedeciba (Uruguay). J.O acknowledges the projects MICINN/MINECO FIS2011-30145-C03-02 and FIS2014-54800-C2-2-P (Spain), and NSF-PHY-1305000 (USA).}

\appendix

\section{Basis for the tensor modes}\label{app:ten-harm}

In this manuscript, we provide an explicit construction of the tensor harmonics $(\tilde G_{ij})_{\vec {\bf n}}$ that we employ to expand in Fourier modes our tensor perturbations. Let us start with the scalar harmonics. Following Refs. \cite{muk-hyb,hybrid,hybrid1}, scalar functions can be decomposed in real Fourier modes within the basis of scalar harmonics given by 
\begin{align}\nonumber
&\tilde Q_{\vec n,+} (\vec\theta)= \sqrt 2\cos\left(\frac{2\pi}{l_0}\vec n\cdot\vec\theta\right),\\
&\tilde Q_{\vec n,-} (\vec\theta)= \sqrt 2\sin\left(\frac{2\pi}{l_0}\vec n\cdot\vec\theta\right),
\end{align}
recalling that $\vec n=(n_1,n_2,n_3)\in\mathbb Z^3$ is any tuple with the first nonvanishing component strictly positive as well as the value $\vec n =  0$ excluded. We use the notation $\vec n\cdot\vec\theta=\sum_in_i\theta_i$. These harmonics are solutions to the equation
\begin{equation}
^0h^{ij}\;^{0}\nabla_i^{0}\nabla_j \tilde Q_{\vec n,\epsilon}= -\omega_n^2\tilde Q_{\vec n,\epsilon},
\end{equation}
with $\epsilon=(+,-)$ and $\omega_n^2=4\pi^2\vec n\cdot\vec n/l_0^{2}$. Here, under a transformation of the form of $\theta_i\to (l_0-\theta_i)$, these scalar real harmonics transform as $\tilde Q_{\vec n,\epsilon}\to\epsilon \tilde Q_{\vec n,\epsilon}$. 

In order to construct an explicit basis for the tensor harmonics $(\tilde G_{ij})_{\vec {\bf n}}$, let us introduce a basis of orthonormal vectors in the tangent bundle of the $T^3$ manifold given by the orthonormal vectors $\hat n$, with
\begin{equation}
\hat n=\frac{\vec n}{\|\vec n\|},
\end{equation}
and the two additional unit vectors $\hat e^{1,\vec{n}}$ and $\hat e^{2,\vec{n}}$ such that $\|\hat e^{a,\vec{n}}\|=1$, $\langle \hat n\cdot \hat e^{a,\vec{n}}\rangle =0$ and $\langle \hat e^{1,\vec{n}}\cdot \hat e^{2,\vec{n}}\rangle =0$, for $a=1,2$. Here, 
\begin{equation}
\langle \hat u\cdot \hat v\rangle=^0h_{ij}u^iv^j
\end{equation}
is the standard inner product with respect to the spatial flat metric $^0h_{ij}$, and $\|\cdot\|$ is the corresponding norm, i.e. $\|v\|=\sqrt{^0h_{ij}v^iv^j}$. It is important to notice that $\hat e^{a,\vec{n}}$ depends on the vector $\vec n$.

With this choice, the tensor harmonics $(\tilde G_{ij})_{\vec {\bf n}}$ take the form
\begin{align}
(\tilde G_{ij})_{\vec {n},\epsilon,+}=\frac{\hat e^{1,\vec{n}}_i\hat e^{1,\vec{n}}_j-\hat e^{2,\vec{n}}_i\hat e^{2,\vec{n}}_j}{\sqrt{2}}\tilde Q_{\vec n,\epsilon},\\
(\tilde G_{ij})_{\vec {n},\epsilon,\times}=\frac{\hat e^{1,\vec{n}}_i\hat e^{2,\vec{n}}_j+\hat e^{2,\vec{n}}_i\hat e^{1,\vec{n}}_j}{\sqrt{2}}\tilde Q_{\vec n,\epsilon},
\end{align}
recalling that ${\vec {\bf n}}=({\vec {n},\epsilon,\tilde \epsilon})$, with $\tilde\epsilon=(+,\times)$. One can easily see that these symmetric tensor harmonics fulfill the transverse, 
\begin{equation}
^0h^{ij}\;^{0}\nabla_i \tilde (\tilde G_{jk})_{\vec {n},\epsilon,\tilde \epsilon}=-\frac{2\pi\epsilon}{l_0}\;^0h^{ij}\;^{0}n_i(\tilde G_{jk})_{\vec {n},-\epsilon,\tilde \epsilon}=0 ,
\end{equation} 
 and traceless $^0h^{ij}(\tilde G_{ij})_{\vec {\bf n}}=0$ conditions, and they satisfy
\begin{equation}
^0h^{ij}\;^{0}\nabla_i^{0}\nabla_j (\tilde G_{kl})_{\vec {\bf n}}= -\omega_n^2(\tilde G_{kl})_{\vec {\bf n}}.
\end{equation}

Besides, they are normalized as
\begin{equation}
\int d^3x \sqrt{^{0}h} (\tilde G_{ij})_{\vec {\bf n}} (\tilde G^{ij})_{\vec {\bf n}'}=\delta_{\vec {\bf n},\vec {\bf n}'}.
\end{equation}
Here, $d^3x \sqrt{^{0}h}$ is the infinitesimal volume element of the spatial manifold.

\bibliographystyle{plain}

\end{document}